\documentstyle[12pt]{article}
\renewcommand{\thesection}{\Roman{section}}
\renewcommand{\theequation}{\thesection-\arabic{equation}}
\bibliography{plain}
\title{\bf Hierarchy of Chaotic maps with an invariant measure and
their coupling.} \vspace{20mm}
\author{M. A.
   Jafarizadeh$^{a,b,c}$\thanks{E-mail:jafarzadeh@ark.tabrizu.ac.ir}
   , S.Behnia$^{d,e,b}$.\\
   $^a${\small Department of Theoretical Physics and Astrophysics,
   Tabriz University, Tabriz 51664, Iran.} \\ $^b${\small Institute
   for Studies in Theoretical Physics and Mathematics, Teheran
   19395-1795, Iran.} \\ $^c${\small Pure and Applied Science
   Research Center, Tabriz 51664, Iran.} \\ $^d${\small Plasma
   Physics Research Center, IAU, Teheran 14835-159, Iran.} \\ $^e $
   {\small Department of Physics, IAU, Urmia, Iran.}} \pagebreak
\begin{document}
\maketitle\vspace{15mm}
\newpage
\begin{abstract}
Hierarchy of one-parameter families of chaotic maps with an
invariant measure have been introduced, where their appropriate
coupling has lead to the generation of  some coupled chaotic maps
with an invariant measure. It is shown that these chaotic maps
(also the coupled maps) do not undergo any period doubling or
period-n-tupling cascade bifurcation to chaos, but they have
either single fixed point attractor at certain values of the
parameters or they are ergodic in the complementary region. Using
the invariant measure or Sinai-Rulle-Bowen measure the
Kolmogrov-Sinai entropy of the chaotic maps (coupled maps) have
been calculated analytically, where the numerical simulations
support the results.
  \\{\bf
  PACs Index: 05.45.Ra, 05.45.Jn, 05.45.Tp }
\end{abstract}
\pagebreak \vspace{7cm}
\section{INTRODUCTION}
In recent years chaos or more properly dynamical systems have
become an important area of research activity. One of the
landmarks in it was the introduction of the concept of
Sinai-Ruelle-Bowen (SRB) measure or natural invariant measure.
This is roughly speaking a measure that is supported on an
attractor and it describes the statistics of the long time
behavior of the orbits for almost every initial condition in the
corresponding basin of attractor. This measure can be obtained by
computing the fixed points of the so called Frobenius-Perron(FP)
operator which can be viewed as a differential-integral operator,
hence, exact determination of invariant measure of dynamical
systems is rather a nontrivial task, such that invariant measure
of few dynamical systems such as one-parameter families of
one-dimensional piece wise linear maps  like the Baker and tent
maps \cite{tasaki,tasaki1,jakob,dorf1} or unimodal maps like
logistic map for certain values of its parameter ( Ulam-Von
Neumann map), one dimensional cubic and quintic maps
\cite{umeno1,umeno2} can be derived analytically. In most of cases
only numerical algorithms, as an example Ulam's method
\cite{froy1,froy,blank} are used for computation of fixed points
of FP operator.\\ Here in this article, we give a hierarchy of
one-parameter families $\Phi_N(\alpha,x)$ of maps of interval
$[0,1]$ with an invariant measure, such that Ulam-Von Neumann map,
one dimensional cubic and quintic maps correspond to three first
maps of this hierarchy for particular value of their parameter.
Also the map appearing in the study of field driven thermostated
systems \cite{dorf1} is topologically conjugate to the first map (
generalized Ulam-Von Neumann) of our hierarchy of family of one
parameter chaotic maps. Then, by an appropriate coupling of these
maps, we generate some coupled chaotic maps with an invariant
measure. We have been able to derived analytically their invariant
measure. Using this measure, we have calculated analytically,
Kolmogorov-Sinai(KS) entropy or equivalently positive Lyapunov
characteristic exponent of these maps, where the numerical
simulations support the analytic calculations. Also it is shown
that these maps have another interesting property, that is, they
bifurcate to chaotic regime without having any period doubling or
period-n-tupling scenario.
\\ The paper organized as follows. In section II the hierarchy of
one parameter families of chaotic maps have been introduced.
Section III is devoted to the appropriate coupling of these maps.
In section IV we talk about their invariant measure. In section V,
using the invariant measure, we give their KS entropy. Finally in
section VI we try to support the analytic calculations with the
numerical simulations. The paper ends with brief conclusion and
five appendices which contains all algebraic calculations and
proofs.
 \section {Hierarchy of one-parameter families of chaotic maps}
One of the  exactly solvable one dimensional maps or the map with
invariant measure is the well known Ulam-Von Neumann map
{\cite{Ulam}}\\ $f(x)=x(1-x)$ (Logistic map
 for $\mu=1$). In reference \cite{umeno1,umeno2} the Ulam-Von Neumann map
 with KS-entropy $\ln(2)$ has been generalized to exactly solvable
 one dimensional cubic and quintic maps with KS-entropy $\ln(3)$ and
 $\ln(5)$, respectively. Here we generalize these maps
to a Hierarchy of one parameter families of  maps with invariant
measure, where the map of Reference \cite{dorf1} is topologically
conjugate to the first map (generalized Ulam-Von Neumann) of this
hierarchy.\\ These one-parameter families of chaotic maps of the
interval $[0,1]$ with an invariant measure can be  defined as the
ratio of polynomials of degree $N$: $$
\Phi_{N}^{(1)}(x,\alpha)=\frac{\alpha^2\left(1+(-1)^N{
}_2F_1(-N,N,\frac{1}{2},x)\right)}
{(\alpha^2+1)+(\alpha^2-1)(-1)^N{ }_2F_1(-N,N,\frac{1}{2},x)}$$
 \begin{equation}
  =\frac{\alpha^2(T_N(\sqrt{x}))^{2}}{1+(\alpha^2-1)(T_N(\sqrt{x})^{2})}\quad ,
  \end{equation}
where $N$ is an integer greater than one. Also $$
_2F_1(-N,N,\frac{1}{2},x)=(-1)^{N}\cos{(2N\arccos\sqrt{x})}=(-1)^{N}T_{2N}(\sqrt{x})
$$ is the hypergeometric polynomials of degree $N$ and $T_{N}$ are
Chebyshev polynomials of type I\cite{wang}, respectively.
Obviously these map the unit interval $[0,1]$ into itself.\\
 $ \Phi_{N}(\alpha,x)$ is (N-1)-model map, that is it has
 $(N-1)$ critical points in unit interval $[0,1]$(see Figures 1,2), since its
 derivative is proportional to derivative of hypergeometric
 polynomial $_2F_1(-N,N,\frac{1}{2},x)$ which is itself a hypergeometric
 polynomial of degree $(N-1)$, hence it has
 $(N-1)$ real roots in unit interval $[0,1]$. Defining Shwarzian
 derivative ${S(\Phi_N(x,\alpha))}$ as\cite{dev}: $$
S\left(\Phi_N(x,\alpha)\right)=\frac{\Phi_{N}^{\prime\prime\prime}(x,\alpha)}{\Phi_{N}^{\prime}(x,\alpha)}-\frac{3}{2}\left({\frac{\Phi_{N}^{\prime\prime}(x,\alpha)}{\Phi_{N}^{\prime}(x,\alpha)}}\right)^2=\left(\frac{\Phi_{N}^{\prime\prime}(x,\alpha)}{\Phi_{N}^{\prime}(x,\alpha)}\right)^{\prime}-\frac{1}{2}\left(\frac{\Phi_{N}^{\prime\prime}(x,\alpha)}{\Phi_{N}^{\prime}(x,\alpha)}\right)^2,
$$
 with a prime denoting a single differential, one can show that:
$$
S\left(\Phi_{N}(x,\alpha)\right)=S\left(_2F_1(-N,N,\frac{1}{2},x)\right)\leq0,
$$
 since $\frac{d}{dx}(_2F_1(-N,N,\frac{1}{2},x))$ can be written
as: $$
\frac{d}{dx}\left(_2F_1(-N,N,\frac{1}{2},x)\right)=A\prod^{N-1}_{i=1}(x-x_i),
$$
 with $0 \leq{x_1}<{x_2}<{x_3}<....<x_{N-1}\leq{1}$, then we
have: $$
S\left(_2F_1(-N,N,\frac{1}{2},x)\right)=\frac{-1}{2}\sum^{N-1}_{J=1}\frac{1}{(x-x_j)^2}-\left(\sum^{N-1}_{J=1}\frac{1}{(x-x_j)}\right)^2<{0}.
$$
 Therefore, the maps $\Phi_{N}(x,\alpha)$ have at most $N+1$
attracting periodic orbits \cite{dev}. As we will show at the end
of this section, these maps have only a single period one stable
fixed points or they are ergodic. \\ Denoting n-composition of
functions $\Phi(x,\alpha)$ by $\Phi^{(n)}$, it is straightforward
to show that the derivative of $\Phi^{(n)}$ at its possible n
periodic points of an n-cycle: $x_{2}=\Phi(x_{1},\alpha),
x_{3}=\Phi(x_{2},\alpha), \cdots , x_{1}=\Phi(x_{n},\alpha)$ is
\begin{equation}
\mid\frac{d}{dx}\Phi^{(n)}\mid=\mid\frac{d}{dx}\overbrace{\left(\Phi^{(1,2)}\circ\Phi^{(1,2)}\circ\cdots\circ\Phi^{(1,2)}(x,\alpha)\right)}^{n}\mid
=\prod_{k=1}^{n}\mid\frac{N}{\alpha}(\alpha^{2}+(1-\alpha^{2})x_{k})\mid,
\end{equation}
since for $x_{k}\in [0,1]$ we have:
$$min(\alpha^{2}+(1-\alpha^{2}x_{k}))=min(1,\alpha^{2}), $$
therefore,
$$min\mid\frac{d}{dx}\Phi^{(n)}\mid=\left(\frac{N}{\alpha}min(1,\alpha^{2})\right)^{n}.$$
Hence the above expression is definitely  greater than one for $
\frac{1}{N}< \alpha < N $, that is: maps do not have any kind of
n-cycle or periodic orbits for $ \frac{1}{N}< \alpha < N $,
actually they are ergodic for this interval of parameter. From
(II-2) it follows the values of $\mid\frac{d}{dx}\Phi^{(n)}\mid$
at n periodic points of the n-cycle belongs to interval [0,1],
varies between ${(N\alpha)}^{n}$ and ${(\frac{N}{\alpha})}^{n}$
for $\alpha<\frac{1}{N}$ and between $(\frac{N}{\alpha})^{n}$ and
$(N\alpha )^{n}$ for $\alpha>N$, respectively.\\ From the
definition of these maps, we see that for odd N, both $x=0$ and
$x=1$ belong to one of the n-cycles, while for even N, only $x=1$
belongs to one of the n-cycles of $\Phi_{N}(x,\alpha)$.\\ For
$\alpha<(\frac{1}{N})$ $(\alpha>{N})$, the formula $(II-2)$
implies that for those cases in which $x=0$ $(x=1)$ belongs to one
of n-cycles we will have $\mid\frac{d}{dx}\Phi^{(n)}\mid<1$, hence
the curve of $\Phi^{(n)}$ starts at $x=0$ $(x=1)$ beneath the
bisector and then crosses it at the next (previous) periodic point
with slope greater than one, since the formula $(II-2)$ implies
that the slope of fixed points increases with the increasing
(decreasing) of $\mid x_{k}\mid$, therefore at all periodic points
of n-cycles except for $x=0$ $(x=1)$ the slope is greater than one
that is they are unstable, this is possible only if $x=0$ $(x=1)$
is the only period one fixed point of these maps.\\ Hence all
n-cycles except for possible period one fixed points $x=0$ and
$x=1$ are unstable, where  for $\alpha\in [0,\frac{1}{N}]$, the
fixed point $x=0$ is stable in maps $\Phi_{N}(x,\alpha)$ (for odd
integer values of N), while for $\alpha\in [N,\infty)$ the $x=1$
is stable fixed point in maps (see Figures 3,4 for $N=2,3$).\\
Below we also introduce their conjugate or isomorphic maps which
will be very useful in derivation of their invariant measure and
calculation of their KS-entropy in the next sections. Conjugacy
means that the invertible map $ h(x)=\frac{1-x}{x} $ maps $
I=[0,1]$ into $ [0,\infty) $ and transform maps
$\Phi_{N}(x,\alpha) $ into $\tilde{\Phi}_{N}(x,\alpha)$ defined
as:
\begin{equation}
\tilde{\Phi}_{N}(x,\alpha)=h\circ\Phi_{N}(x,\alpha)\circ
h^{(-1)}=\frac{1}{\alpha^{2}}\tan^{2}(N\arctan\sqrt{x})
\end{equation}
 As an example we give below some of these maps:
 \begin{equation}
\Phi_{2}(\alpha,x)=\frac{\alpha^{2}(2x-1)^{2}}{4x(1-x)+\alpha^{2}(2x-1)^{2}},
\end{equation}
 \begin{equation}
\Phi_{3}(\alpha,x)=\frac{\alpha^{2}x(4x-3)^{2}}{\alpha^{2}x(4x-3)^{2}+(1-x)(4x-1)^{2}},
\end{equation}
 \begin{equation}
\Phi_{4}(\alpha,x)=\frac{\alpha^{2}(1-8x(1-x))^{2}}{\alpha^2(1-8x(1-x))^{2}+16x(1-x)(1-2x)^{2}},
\end{equation}
 \begin{equation}
\Phi_{5}(\alpha,x)={\frac{{\alpha}^{2}x\left(16\,{x}^{2}-20\,x+5\right
)^{2}}{{\alpha}^ {2}x\left (16\,{x}^{2}-20\,x+5\right )^{2}-x\left
(16\,{x}^{2}-20\,x+5 \right )^{2}+1}}.
\end{equation}
\section {Hierarchy of coupled map}
\setcounter{equation}{0}
 Using the hierarchy of families of one-parameter chaotic maps $(II-1)$, we can
generate new hierarchy of coupled maps with an invariant measure.
Hence, by introducing a new parameter ${\bf g}$ as a coupling
parameter we form coupling among the above mentioned maps, where
they can be coupled through $\bf{\beta}$ and ${\bf \alpha}$
functions defined as:
  \begin{equation}
  \beta(x)=(\sqrt{\beta_{0}}+gx)^{2},
  \end{equation}
  \begin{equation}
   \alpha_{N}(x)=\frac{B_{N}(\frac{1}{\beta(x)})}{A_{N}(\frac{1}{\beta(x)})}\sqrt{\frac{\beta(X)}{\beta(x)}}
  \end{equation}
  with $B_{N}(x)$ and $A_{N}(x)$ defined as:
\begin{equation}
 A(x)=\sum_{k=0}^{[ \frac{N}{2}]}C_{2k}^{N}x^{k},
\end{equation}
\begin{equation}
B(x)=\sum_{k=0}^{[ \frac{N-1}{2}]}C_{2k+1}^{N}x^{k}.
\end{equation}
  Now, the hierarchy of the coupled maps can be defined as:
\begin{equation} \Phi_{N_{1},N_{2}}= \left\{
\begin{array}{l}X_{1}=\Phi_{N_{1}}(x_{1},\alpha_{1}(x_{2}))\\
 X_{2}=\Phi_{N_{1}}(x_{2},\alpha_{2}),
\end{array}\right.
\end{equation}
  Their conjugate or isomorphic maps are defined as:
 \begin{equation}
 \tilde{\Phi}_{N_{1},N_{2}}= \left\{
\begin{array}{l} \tilde{X}_{N_{1}}(x_{1},\alpha_{1}(x_{2}))=h\circ X_{N_{1}}\circ
    h^{(-1)}=\frac{1}{\alpha_{1}^{2}(x_{2})}\tan^{2}(N_{1}\arctan\sqrt{x_{1}})\\
 \tilde{X}_{N_{2}}(x_{2},\alpha_{2})=h\circ X_{N_{2}}\circ
    h^{(-1)}=\frac{1}{\alpha_{2}^{2}}\tan^{2}(N_{2}\arctan\sqrt{x_{2}}).
     \end{array}\right.
 \end{equation}
Below we give some of coupled map as an illustration
\begin{equation}   \Phi_{2,2}=\left\{
\begin{array}{l} X_{1}=\frac{\alpha_{1}^{2}(x_{2})(2x_{1}-1)^{2}}{4x_{1}(1-x_{1})+\alpha_{1}^{2}(x_{2})(2x_{1}-1)^{2}}
\\ \\
 X_{2}=\frac{\alpha_{2}^{2}(2x_{2}-1)^{2}}{4x_{2}(1-x_{2})+\alpha_{2}^{2}(2x_{2}-1)^{2}}
\end{array}\right.
\end{equation}
and
\begin{equation}
  \Phi_{2,3}=\left\{ \begin{array}{l}
  X_{1}=\frac{\alpha_{1}^{2}(x_{2})(2x_{1}-1)^{2}}{4x_{1}(1-x_{1})+\alpha_{1}^{2}(x_{2})(2x_{1}-1)^{2}}
  \\ \\
  X_{2}=\frac{\alpha_{2}^{2}x_{2}(4x_{2}-3)^{2}}{\alpha_{2}^{2}x_{2}(4x_{2}-3)^{2}+(1-x_{2})(4x_{2}-1)^{2}}
  \end{array}\right.
  \end{equation}
  with
  $$
  \alpha_{1}(x_{2})=\frac{2\beta(x_{2})}{1+\beta(x_{2})}\sqrt{\frac{\beta(X_{2})}
  {\beta(x_{2})}},
  $$
$$\beta(x)=(\sqrt{\frac{\alpha_{1}}{2-\alpha_{1}}}+gx)^{2}.$$ One
can show the map $\Phi_{2,3}$, is ergodic, for $0<\alpha_{1}<2$
and $\frac{1}{3}<\alpha_{2}< 3$, while for $2<\alpha_{1}<\infty$
and  $0<\alpha_{2}<\frac{1}{3}$
  ($3<\alpha_{2}<\infty$) it has fixed point at $[0,0]$ ($[0,1]$) (see Fig. 5).
 \begin{equation}
  \Phi_{3,2}=\left\{
  \begin{array}{l}
  X_{1}=\frac{\alpha_{1}^{2}(x_{2})x_{1}(4x_{1}-3)^{2}}{\alpha_{1}^{2}(x_{2})x_{1}(4x_{1}-3)^{2}+(1-x_{1})(4x_{1}-1)^{2}}
  \\ \\
  X_{2}=\frac{\alpha_{2}^{2}(2x_{2}-1)^{2}}{4x_{2}(1-x_{2})+\alpha_{2}^{2}(2x_{2}-1)^{2}}
  \end{array}\right.
  \end{equation}
 Similarly, one can show that the map $\Phi_{3,2}$, is ergodic for
$\frac{1}{3}<\alpha_{1}<3$ and $0<\alpha_{2}<2$, while for
$0<\alpha_{1}<\frac{1}{3}$ ($3<\alpha_{1}<\infty)$ and
$0<\alpha_{2}<2$ ($2<\alpha_{2}<\infty$) it has fixed point at
$[0,0]$ ($[1,0]$) (see Fig. 6).
\begin{equation}
   \Phi_{3,3}=\left\{
  \begin{array}{l}
  X_{1}=\frac{\alpha_{1}^{2}(x_{2})x_{1}(4x_{1}-3)^{2}}{\alpha_{1}^{2}(x_{2})x_{1}(4x_{1}-3)^{2}+(1-x_{1})(4x_{1}-1)^{2}}
  \\ \\
  X_{2}=\frac{\alpha_{2}^{2}x_{2}(4x_{2}-3)^{2}}{\alpha_{2}^{2}x_{2}(4x_{2}-3)^{2}+(1-x_{2})(4x_{2}-1)^{2}}
  \end{array}\right.
  \end{equation}
   with
  $$\alpha_{1}(x_{2})=\frac{3\beta(x_{2})+1}{3+\beta(x_{2})}\sqrt{\frac{\beta(X_{2})}
 {\beta(x_{2})}},$$
$$\beta(x)=(\sqrt{\frac{\alpha_{1}-3}{3\alpha_{1}-1}}+gx)^{2}.$$
Also the map $\Phi_{3,3}$, is ergodic for
$\frac{1}{3}<\alpha_{1}<3$ and $\frac{1}{3}<\alpha_{2}<3$, while
for $0<\alpha_{1}<\frac{1}{3}$ ($3<\alpha_{1}<\infty$ ) and
$0<\alpha_{2}<\frac{1}{3}$  ($3<\alpha_{2}<\infty$ )it has fixed
point at $[0,0]$ ($[1,1]$) (see Fig. 7).
\begin{equation}
  \Phi_{4,3}=\left\{
  \begin{array}{l}
 X_{1}=\frac{\alpha^{2}_{1}(x_{1})(1-8x_{1}(1-x_{1}))^{2}}{\alpha^2_{1}(x_{1})(1-8x_{1}(1-x_{1}))^{2}+16x_{1}(1-x_{1})(1-2x_{2})^{2}}
  \\ \\
  X_{2}=\frac{\alpha^{2}x_{2}(4x_{2}-3)^{2}}{\alpha^{2}x_{2}(4x_{2}-3)^{2}+(1-x_{2})(4x_{2}-1)^{2}}
  \end{array}\right.
  \end{equation}
and
 \begin{equation}
  \Phi_{4,4}=\left\{
  \begin{array}{l}
 X_{1}=\frac{\alpha^{2}_{1}(x_{1})(1-8x_{1}(1-x_{1}))^{2}}{\alpha^2_{1}(x_{1})(1-8x_{1}(1-x_{1}))^{2}+16x_{1}(1-x_{1})(1-2x_{2})^{2}}
  \\ \\
    X_{2}=\frac{\alpha^{2}_{2}(1-8x_{2}(1-x_{2}))^{2}}{\alpha^2(1-8x_{2}(1-x_{2}))^{2}+16x_{2}(1-x_{2})(1-2x_{2})^{2}}
  \end{array}\right.
  \end{equation}
    with
 $$\alpha_{1}(x_{2})=\frac{\beta^{2}(x_{2})+6\beta(x_{2})+1}{\beta(x_{2})(4\beta(x_{2})+4)}
                   \sqrt{\frac{\beta(X_{2})}{\beta(x_{2})}},$$
$$
   \beta(x)=(\sqrt{\frac{4-6\alpha_{1}-4\sqrt{1-2\alpha_{1}+2\alpha_{1}^{2}}}{2(\alpha_{1}-4)}}+gx)^{2}.
  $$
The maps  $\Phi_{2,2}$, $\Phi_{4,3}$ and $\Phi_{4,4}$ are also
either ergodic or they only single fixed points at $[0,0]$ or
$[0,1]$.
 \section{Invariant measure}
 \setcounter{equation}{0}
 Dynamical systems, even apparently simple dynamical systems which are described by maps of an interval can
display a rich variety of different asymptotic behavior. On
measure theoretical level these type of behavior are described by
SRB \cite{sinai,dorf2} or invariant measure describing
statistically stationary states of the system. The probability
measure ${\bf \mu}$ on $[0,1]$ is called an SRB or invariant
measure of the map $y=\Phi_{N}(x,\alpha)$ given in $(II-1)$, if it
is $\Phi_N(x,\alpha)$-invariant and absolutely continuous with
respect to Lebesgue measure. For deterministic system such as
$\Phi_{N}(x,\alpha)$-map, the $\Phi_N(x,\alpha)$-invariance means
that, its invariant measure $\mu(x)$ fulfills the following formal
FP integral equation
$$\mu(y)=\int_{0}^{1}\delta(y-\Phi_N(x,\alpha))\mu(x)dx.$$
 This is equivalent to:
\begin{equation}
\mu(y)=\sum_{x\in\Phi_{N}^{-1}(y,\alpha)}\mu(x)\frac{dx}{dy}\quad,
\end{equation}
defining the action of standard FP operator for the map
$\Phi_N(x,\alpha)$ over a function as:
\begin{equation}
P_{\Phi_{N}}f(y)=\sum_{x\in
\Phi_{N}^{-1}(y,\alpha)}f(x)\frac{dx}{dy}\quad.
\end{equation}
We see that, the invariant measure $\mu(x)$ is actually the
eigenstate of the FP operator $P_{\Phi_N}$ corresponding to
largest eigenvalue 1.\\ As it is proved in Appendix A, the
invariant measure $\mu_{\Phi_{N}(x,\alpha)}(x,\beta)$ has the
following form:
\begin{equation}
\mu{(x)}=\frac{1}{\pi}\frac{\sqrt{\beta}}{\sqrt{x(1-x)}(\beta+(1-\beta)x)}
\end{equation}
with $\beta>0$, is the invariant measure of the maps
$\Phi_{N}(x,\alpha)$ provided that, we choose the parameter
$\alpha$ in the following form :
\begin{equation}
\alpha=\frac{\sum{^{[\frac{(N-1)}{2}]}_{k=0}
C^{N}_{2k+1}\beta^{-k}}}{\sum{^{[\frac{N}{2}]}_{k=0}
C^{N}_{2k}\beta^{-k}}},
\end{equation}
where the symbol $[\quad]$ means greatest integer part.\\
 As we see, the above measure is defined only for $\beta > 0$, hence from the
 relations $(IV-3)$, it follows that the maps $\Phi_N(x,\alpha)$ have
invariant  measure only for $\alpha\in{(\frac{1}{N}, N)}$ for odd
value of N, and for even value $\alpha\in{(N, \infty)}$. For other
values of $\alpha$ these maps have single attractive fixed points,
which is the same as the prediction of the previous section.
\\ Similarly the probability measure ${\bf \mu}$ for coupled chaotic maps $\Phi_{N_{1},N_{2}}$
given in $(III-4)$ fulfills the following formal FP integral
equation $$\mu(y_{1},y_{2})=\int_{0}^{1}dx_{1}\int_{0}^{1}dx_{2}
                      \delta(y_{1}-X_{1}(x_{1},x_{2}))
                       \delta(y_{2}-X_{2}(x_{1},x_{2}))\mu(x_{1},x_{2}),$$
 which is equivalent to:
 \begin{equation}
  \mu(y_{1},y_{2})=\sum_{x \varepsilon \Phi^{-1}_{N_{1},N_{2}}}{\mid J(x_{1},x_{2})
  \mid}\mu(x_{1},x_{2}),
 \end{equation}
 where $J(x_{1},x_{2})$ is the Jacobian of the transformation
 $\Phi_{N_{1},N_{2}}$ which is equal to
  $$ J(x_{1},x_{2})=\frac{\partial (x_{1},x_{2})}{\partial (y_{1},y_{2})},$$
 again defining the action of standard FP operator for the map
$\Phi_{N_{1},N_{2}}$ over a function as:
\begin{equation}
P_{\Phi_{N_{1},N_{2}}}f(y)=\sum_{x\in
\Phi_{N_{1},N_{2}}^{-1}}f(x)J(x_{1},x_{2}).
\end{equation}
We see that, the invariant measure $\mu(x_{1},x_{2})$ is also the
eigenstate of the FP operator $P_{\Phi_{N_{1},N_{2}}}$
corresponding to largest eigenvalue 1.\\ As it is proved in
Appendix B the invariant measure
$\mu_{\Phi_{N_{1},N_{2}}}(x_{1},x_{2})$ has the following form:
\begin{equation}
\frac{1}{\pi}\frac{\sqrt{\beta_{2}}}{\sqrt{x_{2}(1-x_{2})}(\beta_{2}+(1-\beta_{2})x_{2})}
\times\frac{1}{\pi}\frac{\sqrt{\beta_{1}(x_{2})}}{\sqrt{x_{1}(1-x_{1})}(\beta_{1}(x_{2})+(1-\beta_{1}(x_{2}))x_{1})},
\end{equation}
with $\beta_{2}>0$ and $\beta_{1}(x)>0$ given in $(III-1)$.
\section{Kolmogrov-Sinai entropy}
\setcounter{equation}{0}
 KS-entropy or metric entropy \cite{sinai}
measures how chaotic a dynamical system is and it is proportional
to the rate at which information about the state of dynamical
system is lost in the course of time or iteration. Therefore, it
can also be defined as the average rate of information loss for a
discrete measurable dynamical system $(\Phi_{N}(x,\alpha),\mu)$.
By introducing a partition $\alpha={A_c} (n_1,.....n_{\gamma})$ of
the interval $[0,1]$ into individual laps $A_i$ one can define the
usual entropy associated with the partition by:
$$H(\mu,\gamma)=-\sum^{n(\gamma)}_{i=1}m(A_c)\ln{m(A_c)},$$ where
$m(A_c)=\int{_{n\in{A_i}}\mu(x)dx}$ is the invariant measure of $
A_i$. By defining n-th refining $\gamma(n)$ of  $ \gamma$ as:
$$\gamma^{n}=\bigcup^{n-1}_{k=0}(\Phi_{N}(x,\alpha))^{-(k)}(\gamma),$$
then entropy per unit step of refining is defined by :
$$h(\mu,\Phi_{N}(x,\alpha),\gamma)=\lim{_{n\rightarrow{\infty}}}(\frac{1}{n}H(\mu,\gamma))$$
Now, if the size of individual laps of $\gamma(N)$ tends to zero
as n increases, then the above entropy is known as KS-entropy,
that is:
 $$h(\mu,\Phi_{N}(x,\alpha)) =h(\mu,\Phi_{N}(x,\alpha), \gamma).$$
 KS-entropy, which is a quantitative measure of the rate of information
 loss with the refining, may also be written as:
\begin{equation}
h(\mu,\Phi_{N}(x,\alpha))
=\int{\mu(x)dx}\ln{\mid\frac{d}{dx}\Phi_{N}(x,\alpha)\mid},
\end{equation}
which is also a statistical mechanical expression for the Lyapunov
characteristic exponent, that is, the mean divergence rate of two
nearby orbits. The measurable dynamical system $(\Phi,\mu)$ is
chaotic for $h>0$ and predictive for $h=0$. As It is proved in
Appendix C, the $h(\mu,\Phi_{N}(x,\alpha))$ has the following form
\begin{equation}
 h(\mu,\Phi_{N}^{(2)}(x,\alpha))=\ln\left(\frac{N(1+\beta+2\sqrt{\beta})^{N-1}}{(\sum_{k=0}^{[ \frac{N}{2}]}C_{2k}^{N}\beta^{k})(\sum_{k=0}^{[
 \frac{N-1}{2}]}C_{2k+1}^{N}\beta^{k})}\right).
\end{equation}
Similarly the KS-entropy of coupled chaotic maps can be written as
 \begin{equation}
  h(\mu,\Phi_{N_{1},N_{2}})=
  \int_{0}^{\infty}dx_{1}
  \int_{0}^{\infty}{dx_{2}\tilde{\mu}(x_{2},x_{2})}
  \ln{\mid\frac{ \partial (\tilde{X}_{N_{1}},\tilde{X}_{N_{2}})}{\partial (x_{1},x_{2})}
  \mid}.
  \end{equation}
In Appendix D we have calculated KS-entropy of coupled chaotic
maps $\Phi_{N_1,N_{2}}$ for $N_1=2,3,4$ and $5$, where below we
give the results.
\newpage
  $$
    h(\mu,\Phi_{2,N_{2}})=
    \ln\left(\frac{N_{2}(1+\beta_{2}+2\sqrt{\beta_{2}})^{N_{2}-1}}
     {(\sum_{k=0}^{[
     \frac{N_{2}}{2}]}C_{2k}^{N_{2}}\beta_{2}^{k})(\sum_{k=0}^{[
     \frac{N_{2}-1}{2}]}C_{2k+1}^{N_{2}}\beta_{2}^{k})}\right)$$
 \begin{equation}
   +\ln{\frac{\left(\sqrt{1+\sqrt{\beta_{1_{0}}}}+\sqrt{\frac{g}{\beta_{2}}}\right)^{4}}{\left(\sqrt{1+\beta_{1_{0}}}+\frac{g}{\beta_{2}}+
               \sqrt{\frac{2g\sqrt{\beta_{1_{0}}}}{\beta_{2}}
                     +\frac{2g}{\beta_{2}}\sqrt{1+\beta_{1_{0}}}}\right)^{2}}},
 \end{equation}
 $$
     h(\mu,\Phi_{3,N_{2}}) =
     \ln\left(\frac{N_{2}(1+\beta_{2}+2\sqrt{\beta_{2}})^{N_{2}-1}}
     {(\sum_{k=0}^{[
     \frac{N_{2}}{2}]}C_{2k}^{N_{2}}\beta_{2}^{k})(\sum_{k=0}^{[
     \frac{N_{2}-1}{2}]}C_{2k+1}^{N_{2}}\beta_{2}^{k})}\right)$$
 $$+\ln\left[{3(\sqrt{1+\sqrt{\beta_{1_{0}}}}+\sqrt{\frac{g}{\beta_{2}}})^{8}/}\right.$$
   $$\left({(\sqrt{3+\beta_{1_{0}}}
         +\frac{g}{\beta_{2}}
         +\sqrt{\frac{2g\sqrt{\beta_{1_{0}}}}{\beta_{2}}
                      +\frac{2g}{\beta_{2}}\sqrt{(3+\beta_{1_{0}})}})}\right)^{2}$$
 \begin{equation}
  \left. \times\left({(\sqrt{1+3\beta_{1_{0}}} +\frac{\sqrt{3}g}{\beta_{2}}
         +\sqrt{\frac{6g\sqrt{\beta_{1_{0}}}}{\beta_{2}}
         +\frac{2g}{\beta_{2}}\sqrt{3(1+3\beta_{1_{0}})}}}\right)^{2}\right],
 \end{equation}
 $$
     h(\mu,\Phi_{4,N_{2}})=
     \ln\left(\frac{N_{2}(1+\beta_{2}+2\sqrt{\beta_{2}})^{N_{2}-1}}
     {(\sum_{k=0}^{[
     \frac{N_{2}}{2}]}C_{2k}^{N_{2}}\beta_{2}^{k})(\sum_{k=0}^{[
     \frac{N_{2}-1}{2}]}C_{2k+1}^{N_{2}}\beta_{2}^{k})}\right)$$
 $$+\ln\left[{4(\sqrt{1+\sqrt{\beta_{1_{0}}}}+\sqrt{\frac{g}{\beta_{2}}})^{12}/}\right.$$
   $$\left({(\sqrt{3-2\sqrt{2}+\beta_{1_{0}}}
         +\frac{g}{\beta_{2}}
         +\sqrt{\frac{2g\sqrt{\beta_{1_{0}}}}{\beta_{2}}
                      +\frac{2g}{\beta_{2}}\sqrt{(3-2\sqrt{2}+\beta_{1_{0}})}})}\right)^{2}$$
$$
   \times\left({(\sqrt{3+2\sqrt{2}+\beta_{1_{0}}} +\frac{g}{\beta_{2}}
         +\sqrt{\frac{2g\sqrt{\beta_{1_{0}}}}{\beta_{2}}
         +\frac{2g}{\beta_{2}}\sqrt{(3+2\sqrt{2}+\beta_{1_{0}})}})}\right)^{2}
 $$
 \begin{equation}
  \left.\times\left({(\sqrt{4+4\beta_{1_{0}}} +\frac{4g}{\beta_{2}}
         +\sqrt{\frac{8g\sqrt{\beta_{1_{0}}}}{\beta_{2}}
         +\frac{2g}{\beta_{2}}\sqrt{(4+4\beta_{1_{0}})}})}\right)^{2}\right],
 \end{equation}
 and
 \newpage
  $$
     h(\mu,\Phi_{5,N_{2}})=
     \ln\left(\frac{N_{2}(1+\beta_{2}+2\sqrt{\beta_{2}})^{N_{2}-1}}
     {(\sum_{k=0}^{[
     \frac{N_{2}}{2}]}C_{2k}^{N_{2}}\beta_{2}^{k})(\sum_{k=0}^{[
     \frac{N_{2}-1}{2}]}C_{2k+1}^{N_{2}}\beta_{2}^{k})}\right)$$
 $$+\ln\left[{5(\sqrt{1+\sqrt{\beta_{1_{0}}}}+\sqrt{\frac{g}{\beta_{2}}})^{16}/}\right.$$
   $$\left({(\sqrt{5-2\sqrt{2}+\beta_{1_{0}}}
         +\frac{g}{\beta_{2}}
         +\sqrt{\frac{2g\sqrt{\beta_{1_{0}}}}{\beta_{2}}
                      +\frac{2g}{\beta_{2}}\sqrt{(5-2\sqrt{2}+\beta_{1_{0}})}})}\right)^{2}$$
$$
   \times\left({(\sqrt{5+2\sqrt{2}+\beta_{1_{0}}} +\frac{g}{\beta_{2}}
         +\sqrt{\frac{2g\sqrt{\beta_{1_{0}}}}{\beta_{2}}
         +\frac{2g}{\beta_{2}}\sqrt{(5+2\sqrt{2}+\beta_{1_{0}})}})}\right)^{2}
 $$
 $$
   \times\left({(\sqrt{1+\frac{2}{5}\sqrt{2}+\beta_{1_{0}}} +\frac{g}{\beta_{2}}
         +\sqrt{\frac{2g\sqrt{\beta_{1_{0}}}}{\beta_{2}}
         +\frac{2g}{\beta_{2}}\sqrt{(1+\frac{2}{5}\sqrt{2}+\beta_{1_{0}})}})}\right)^{2}
 $$
 \begin{equation}
  \left.\times\left({(\sqrt{1-\frac{2}{5}\sqrt{2}+\beta_{1_{0}}} +\frac{g}{\beta_{2}}
         +\sqrt{\frac{2g\sqrt{\beta_{1_{0}}}}{\beta_{2}}
         +\frac{2g}{\beta_{2}}\sqrt{(1-\frac{2}{5}\sqrt{2}+\beta_{1_{0}})}})}\right)^{2}\right].
 \end{equation}

\section{Simulation}
 \setcounter{equation}{0}
 Here in this section we try to calculate Lyapunov characteristic exponent of
maps $(II-3,4)$ and $(III-6,7)$ in order to investigate these maps
numerically. In fact, Lyapunov characteristic exponent is the
characteristic exponent of the rate of average magnification of
the neighborhood of an arbitrary point
$\vec{r}_{0}=(x_{1_{0}},x_{2_{0}})$ and it is denoted by $
\Lambda(\vec{r}_{0})$ which is written as\cite{dorf2}:
$$\Lambda(\vec{r}_{0})=\lim_{n \rightarrow \infty}
 \ln{\mid\frac{\partial ((\overbrace{\Phi \circ \Phi \circ ....\circ
\Phi}^{k})_{1} ,(\overbrace{\Phi \circ \Phi \circ ....\circ
\Phi}^{k})_{2})}
 {\partial (x_{1_{0}},x_{2_{0}})}\mid}$$
\begin{equation}
\Lambda(\vec{r}_{0})=\lim_{n \rightarrow \infty}\sum_{k=0}^{n-1}
\ln{\mid \frac{\partial X_{1}(x_{k},\alpha_{2})}{\partial
x_{1_{0}}}.\frac{\partial X_{2}(x_{k},\alpha_{1}) }{\partial
x_{2_{0}}}\mid}
\end{equation}
where $ x_{k}=\overbrace{\Phi \circ \Phi \circ ....\circ
\Phi}^{k}$ . It is obvious that $\Lambda(\vec{r}_{0})<0 $ for an
attractor, $\Lambda(\vec{r}_{0})>0$ for a repeller and
$\Lambda(\vec{r}_{0})=0$ for marginal situation. Also the Liapunov
number is independent of initial point $\vec{r}_{0}$, provided
that the motion inside the invariant manifold is ergodic, thus
$\Lambda(\vec{r}_{0})$ characterizes the invariant manifold of
$\Phi$ as a whole. For the values of parameter $\alpha$, such that
the map $\Phi$ be measurable, Birkohf ergodic \cite{keller}theorem
implies the equality of KS-entropy and Liapunov number, that is:
\begin{equation}
h(\mu,\Phi)=\Lambda(\vec{r}_{0},\Phi),
\end{equation}
Comparison of analytically calculated KS-entropy of maps (see
 Figures 8-11) with the corresponding
Lyapunov characteristic exponent obtained by the simulation,
indicates that in chaotic region, these maps are ergodic as
Birkohf ergodic theorem predicts. In non chaotic region of the
parameter, Lyapunov characteristic exponent is negative definite,
since in this region we have only single period fixed points
without bifurcation. Therefore, combining the analytic discussion
of section V with the numerical simulation we deduce that, these
maps are ergodic in certain values of their parameters as
explained above and in complementary interval of parameters they
have only a single period one attractive fixed point, such that in
contrary to the most of usual one-dimensional one-parameter family
of maps they have only bifurcation from a period one attractive
fixed point to chaotic state or vise versa.
 A comparison of these results with those of obtained in section V,
 verified the accuracy of implemented computation methods.
\section{Conclusion}
 \setcounter{equation}{0}
  We have given hierarchy of exactly
solvable one-parameter families of one-dimensional chaotic maps
 with an invariant measure, that is measurable dynamical system
with an interesting property of being either chaotic (proper to
say ergodic ) or having  stable period one fixed point and they
bifurcate from a stable single periodic state to chaotic one and
vice-versa without having usual period doubling or
period-n-tupling scenario. Also by coupling these maps in an
appropriate method, we have obtained some coupled chaotic maps
with invariant measure, where they also bifurcate without
period-n-tupling scenario to chaos. Since the invariant measure
plays an important roles in analytic calculation of important
equations such as KS-entropy, therefore it would be interesting to
find the other measurable one parameter maps, many parameters,
coupled and  higher dimensional maps with invariant measure, which
are under investigation.
\newpage
\renewcommand{\thesection}{A}
\renewcommand{\theequation}{\thesection-\arabic{equation}}
\setcounter{equation}{0}
 {\large \appendix{\bf Appendix A}}: Derivation of measure of
 one-parameter families of chaotic maps
 \\ In order to prove that measure $(IV-3)$ satisfies equation
 $(IV-1)$, it is rather convenient to consider the conjugate map.
\begin{equation}
\tilde{\Phi}_{N}(x,\alpha)=\frac{1}{\alpha^2}\tan^{2}(N\arctan\sqrt{x})
\end{equation}
with measure $\tilde{\mu}_{\tilde{\Phi}_{N}}$ related to the
measure $\mu_{\Phi_{N}}$ with the following relation:
$$\tilde{\mu}_{\tilde{\Phi}_{N}}(x)=\frac{1}{(1+x)^2}\mu_{\Phi_{N}}(\frac{1}{1+x}).$$
Denoting $\tilde{\Phi}_{N}(x,\alpha)$ on the left hand side of
$(A-1)$ by $y$ and inverting it, we get :
\begin{equation}
x_k=\tan^2(\frac{1}{N}\arctan\sqrt{y\alpha^2}+\frac{k\pi}{N})\quad\quad
k=1,..,N.
\end{equation}
 Then, taking derivative of $x_k$ with respect to $y$, we obtain:
\begin{equation}
\mid\frac{dx_{k}}{dy}\mid=\frac{\alpha}{N}\sqrt{x_{k}}(1+x_k)\frac{1}{\sqrt{y}(1+\alpha^2y)}\quad.
\end{equation}
Substituting the above result in equation $(IV-1)$, we get:
\begin{equation}
\tilde{\mu}_{\tilde{\Phi}_{N}}(y)\sqrt{y}(1+\alpha^2y)=\frac{\alpha}{N}\sum_{k}\sqrt{x_{k}}(1+x_{k})\tilde{\mu}_{\tilde{\Phi}_{N}}(x_k)\quad,
\end{equation}
Now, by considering the following anatz for the invariant measure
$\tilde{\mu}_{\tilde{\Phi}_{N}}(y)$:
\begin{equation}
\tilde{\mu}_{\tilde{\Phi}_{N}}(y)=\frac{\sqrt{\beta}}{\sqrt{y}(1+\beta
y)},
\end{equation}
the above equation reduces to: $$\frac{1+\alpha^2y}{1+\beta
y}=\frac{\alpha}{N}\sum_{k=1}^{N}\left(\frac{1+x_{k}}{1+\beta
x_{k}}\right)$$ which can be written as:
\begin{equation}
\frac{1+\alpha^2y}{1+\beta
y}=\frac{\alpha}{\beta}+\left(\frac{\beta-1}{\beta^{2}}\right)\frac{\partial}{\partial\beta^{-1}}(\ln(\Pi_{k=1}^{N}(\beta^{-1}+x_{k}))).
\end{equation}
To evaluate the second term in the right hand side of above
formulas we can write the equation in the following form: $$
0=\alpha^{2}y\cos^{2}(N\arctan\sqrt{x})-\sin^{2}(N\arctan\sqrt{x})
$$ $$
={\frac{(-1)^{N}}{(1+x)^{N}}}\left(\alpha^{2}y(\sum_{k=0}^{[\frac{N}{2}]}C_{2k}^{N}(-1)^{N}x^{k})^{2}-x(\sum_{k=0}^{[\frac{N-1}{2}]}C_{2k+1}^{N}(-1)^{N}x^{k})^{2}\right),
$$
 $$=\frac{\mbox{constant}}{(1+x)^{N}}\prod_{k=1}^{N}(x-x_{k}),
$$
 where $x_{k}$ are the roots of equation $(A-1)$ and they are given by
formula $(A-2)$. Therefore, we have:
$$\frac{\partial}{\partial\beta^{-1}}\ln\left(\prod_{k=1}^{N}(\beta^{-1}+x_{k})\right)$$
 $$=\frac{\partial}{\partial\beta^{-1}}\ln\left[(1-\beta^{-1})^{N}(\alpha^{2}y\cos^{2}(N\arctan\sqrt{-\beta^{-1}})-\sin^{2}(N\arctan\sqrt{-\beta^{-1}}))\right]$$
\begin{equation}
=-\frac{N\beta}{\beta-1}+\frac{\beta
N(1+\alpha^2y)A(\frac{1}{\beta})
}{(A(\frac{1}{\beta}))^{2}\beta^{2}y+(B( \frac{1}{\beta}))^2},
\end{equation}
In deriving of above formulas we have used the following
identities
$$\cos(N\arctan\sqrt{x})=\frac{A(-x)}{(1+x)^{\frac{N}{2}}},$$
\begin{equation}
\sin(N\arctan\sqrt{x})=\sqrt{x}\frac{B(-x)}{(1+x)^{\frac{N}{2}}},
\end{equation}
 inserting the results $(A-7)$ in $(A-6)$, we get:
 $$\frac{1+\alpha^{2}y}{1+\beta
y}=\frac{1+\alpha^{2}y}{\left( \frac{B( \frac{1}{\beta})}{\alpha
A( \frac{1}{\beta})}+\beta( \frac{\alpha A( \frac{1}{\beta})}{B(
\frac{1}{\beta})})y\right)}.$$
 Hence to get the final result we have to choose the parameter
 $\alpha$ as:
$$\alpha =\frac{B( \frac{1}{\beta})}{A(\frac{1}{\beta})}.$$
\newpage
\renewcommand{\thesection}{B}
\renewcommand{\theequation}{\thesection-\arabic{equation}}
\setcounter{equation}{0}
 {\large \appendix{\bf Appendix B}}: Derivation of measure of coupled chaotic
 maps.
\\ In order to prove that measure $(IV-7)$ satisfies equation
$(IV-5)$, it is rather convenient to consider the conjugate map.
\\ Now, denoting $\tilde{X}_{1}$ by $y_{1}$ and
$\tilde{X}_{2}$ by $y_{2}$ and inverting the conjugate map
$\tilde{\Phi}_{N_{1},N_{2}}$ we get:
 \begin{equation}
  \left\{
 \begin{array}{l}
 x_{k_{1}k_{2}}=\tan^{2}{(\frac{1}{N_{1}}\arctan{\sqrt{y_{1}\alpha_{1}^{2}(x_{2})}}+\frac{k_{1}\pi}{
   N_{1}})}
 \\ \\
 x_{k_{2}}=\tan^{2}{(\frac{1}{N_{2}}\arctan{\sqrt{y_{2}\alpha_{2}^{2}}}+\frac{k_{2}\pi}{ N_{2}})},
 \end{array}\right.
 \end{equation}
then taking the derivative of $(B-1)$ with respect to $y_{1}$,
$y_{2}$:
 \begin{equation}
 \left\{
 \begin{array}{l}
 \frac{\partial x_{k_{1}k_{2}}}{\partial y_{1}}=\frac{g_{1}(x_{k_{2}})}{N_{1}}.
  \frac{\sqrt{x_{k_{1}k_{2}}}
        (1+x_{k_{1}k_{2}})}{\sqrt{y_{1}}(1+g_{1}^{2}(x_{k_{2}})y_{1})}
 \\ \\
 \frac{\partial x_{k_{2}}}{\partial y_{2}}=\frac{\alpha_{2}}{N_{2}}.\frac{(
 1+x_{k_{2}})}{(1+\alpha_{2}^{2}y_{2})}.\frac{\sqrt{x_{k_{2}}}}{\sqrt{y_{2}}},
 \end{array}\right.
 \end{equation}
 and substituting the above results in equation $(IV-1)$, we obtain:
$$\tilde{\mu}(y_{1},y_{2})=\sum_{k_{1},k_{2}}\mid \frac{\partial
x_{k_{2}}}{\partial y_{2}}
                         .\frac{\partial x_{k_{1}k_{2}}}{\partial
                         y_{1}}\mid
\tilde{\mu}(x_{k_{1}k_{2}},x_{k_{2}})$$
\begin{equation}
 =\sum_{k_{1},k_{2}}
 \frac{g_{1}(x_{k_{2}})}{N_{1}}
 .\frac{\sqrt{x_{k_{1}k_{2}}}
        (1+x_{k_{1}k_{2}})}{\sqrt{y_{1}}(1+g_{1}^{2}(x_{k_{2}})y_{1})}
   .\frac{\alpha_{2}}{N_{2}}.\frac{(
 1+x_{k_{2}})}{(1+\alpha_{2}^{2}y_{2})}.\frac{\sqrt{x_{k_{2}}}}{\sqrt{y_{2}}}\tilde{\mu}(x_{k_{1}k_{2}},x_{k_{2}}).
 \end{equation}
Now, considering the following anzatz for invariant measure
  $\tilde{\mu}_{\tilde{\Phi}_{N_{1},N_{2}}}$
 \begin{equation}
 \tilde{\mu}(x_1,x_{2})=\frac{1}{\pi}.\frac{\gamma(x_{2})}
   {\sqrt{x_1}(1+\beta(x_2)x_1)},
 \end{equation}
 we come to the following equation:
\begin{equation}
 \tilde{\mu}(y_{1},y_{2})
  =\frac{1}{\pi}\sum_{k_{2}}\frac{\partial x_{k_{2}}}{\partial y_{2}}
   .\frac{1
   }{\sqrt{y_{1}}}
       .\frac{\gamma(x_{k_{2}})}
            {(1+g_{1}^{2}(x_{k_{2}})y_{1})}
           \times
           \sum_{k_{1}}\frac{g_{1}(x_{k_{2}})}{N_{1}}
           .\frac{(1+x_{k_{1}k_{2}})}
           {(1+\beta_{1}(x_{k_{2}})x_{k_{1}k_{2}})},
 \end{equation}
  using the relation $(III-3)$ in the last part of above relation, it reads
  \newpage
 $$
 \sum_{k_{1}}\frac{g_{1}(x_{k_{2}})}{N_{1}}
                .\frac{(1+x_{k_{1}k_{2}})}
                     {(1+\beta_{1}(x_{k_{2}})x_{k_{1}k_{2}})}=
 \frac{g_{1}(x_{k_{2}})A_{N_{1}}(\beta_{1}^{-1}(x_{k_{2}}))(1+g_{1}^{2}(x_{k_{2}})y_{1})}
                        {B_{N_{1}}(\beta_{1}^{-1}(x_{k_{2}}))}$$
                        $$\times\frac{1}{(1+\beta_{1}(x_{k_{2}})
                        (\frac{g_{1}(x_{k_{2}})A_{N_{1}}(\beta_{1}^{-1}(x_{k_{2}}))}
                        {B_{N_{1}}(\beta_{1}^{-1}(x_{k_{2}}))}))},
$$
 therefore, the relation $(B-5)$ reduces to:
 $$
 \tilde{\mu}(y_1,y_2)=\frac{1}{\pi}\sum_{k_{2}}
 \left( \frac{\gamma(x_{k_{2}})}{\sqrt{y_{1}}}
   .\frac{g_{1}(x_{k_{2}})A_{N_{1}}(\beta_{1}^{-1}(x_{k_{2}}))}
                        {B_{N_{1}}(\beta_{1}^{-1}(x_{k_{2}}))}\right.$$
 \begin{equation}
                       \left. \times\frac{1}{(1+y_{1}\beta_{1}(x_{k_{2}})
                        (\frac{g_{1}(x_{k_{2}})A_{N_{1}}(\beta_{1}^{-1}(x_{k_{2}}))}
                        {B_{N_{1}}(\beta_{1}^{-1}(x_{k_{2}}))})^{2})}.\frac{\partial
                        x_{k_{2}}}{\partial y_{2}}\right).
 \end{equation}
On the otherhand, we know that the measure $\tilde{\mu}(y_1,y_2)$
is
\begin{equation}
 \tilde{\mu}(y_1,y_2)=\frac{1}{\pi}.\frac{\gamma(y_{2})}{(1+\beta(y_{2})y_{1})\sqrt{y_{1}}},
 \end{equation}
which is possible if $\beta_{1}(x_{k_{2}})$ and $\beta_{1}(y_{2})$
to be relate with
  \begin{equation}
  \beta_{1}(y_{2})=\left(\frac{g_{1}(x_{k_{2}})A_{N_{1}}(\beta_{1}^{-1}(x_{k_{2}}))}
         {B_{N_{1}}(\beta_{1}^{-1}(x_{k_{2}}))}\right)\beta_{1}(x_{k_{2}})
  \end{equation}
 and $\gamma(y_{2})$ can be given in terms of $\gamma(x_{k})$ as:
 $$\gamma(y_{2})=\sum_{k_{2}}\left(\frac{\sqrt{\beta_{1}(y_{2})}}
      {\sqrt{\beta_{1}(x_{k_{2}})}}.\frac{\partial x_{k_{2}}}{\partial y_{2}}\right)\gamma(x_{k_{2}}),$$
 or it can be written as
 $$\frac{\gamma (y_{2})}{\sqrt{\beta_{1}(y_{2})}}=\sum_{k_{2}}\frac{\partial x_{k_{2}}}{\partial y_{2}}
 .\frac{\gamma (x_{k_{2}})}{\sqrt{\beta_{1}(x_{k_{2}})}},$$
Therefore, $\frac{\gamma (y_{2})}{\sqrt{\beta_{1}(y_{2})}}$
satisfies the PF equation of map $\Phi_{N}$, hence it is
proportional to its invariant measure, that is
 \begin{equation}
 \frac{\gamma
 (y_{2})}{\sqrt{\beta_{1}(y_{2}})}=\frac{1}{\pi}.\frac{\sqrt{\beta_{2}}}{(1+\beta_{2}y_{2})}.
 \end{equation}
 \newpage
 \newpage
\renewcommand{\thesection}{C}
\renewcommand{\theequation}{\thesection-\arabic{equation}}
\setcounter{equation}{0}
 {\large \appendix{\bf Appendix C}}: Derivation of entropy of one-parameter
 families of chaotic maps
 \\ In order to calculate the KS-entropy of the maps
$\Phi_{N}(x,\alpha)$, it is rather convenient to consider their
conjugate maps given by $(II-2)$, since it can be shown that
KS-entropy is a kind of topological invariant, that is, it is
preserved under conjugacy map. Hence we have: $$ h(\mu,\Phi)
=h(\tilde{\mu},\tilde{\Phi}).$$ The integral $(V-1)$ can be
written as
 $$h(\mu,\Phi_{N}(x,\alpha))=\int_{0}^{\infty}{\tilde{\mu} (x)dx}
   \ln{\mid \frac{d}{dx}(
   \frac{1}{\alpha^{2}}.\tan^{2}(N\arctan\sqrt{x}))\mid}$$
we have
    $$h(\mu,\Phi_{N}(x,\alpha))=\int_{0}^{\infty}{\tilde{\mu}(x)dx}
      \ln{\left( \mid \frac{N}{\alpha^{2}}.\frac{1}{\sqrt{x}(1+x)}
      \frac{\sin N(\arctan\sqrt{x})}{\cos^{3}N(\arctan\sqrt{x})}\mid\right)},$$
using the relation $(A-8)$, we get
\begin{equation}
h(\mu,\Phi_{N}(x,\alpha))
=\frac{1}{\pi}\int_{0}^{\infty}\frac{\sqrt{\beta}dx}{\sqrt{x}(1+\beta
x)}\ln\left(\frac{N}{\alpha^{2}}\mid\frac{(1+x)^{N-1}B(-x)}{(A(-x))^{3}
}\mid\right).
\end{equation}
 We see that polynomials appearing in the numerator ( denominator ) of
integrand appearing on the right hand side of equation $(C-1)$,
have $\frac{[N-1]}{2}$ $(\frac{[N]}{2})$ simple roots, denoted by
 $
  x_{k}^{B}\quad k=1,...,[\frac{N-1}{2}]\quad$
$(x_{k}^{A}\quad k=1,...,[\frac{N}{2}])
$
 in the interval $[0,\infty)$.
 Hence, we can write the above formula in the following form:
 $$h(\mu,\Phi_{N}(x,\alpha))
=\frac{1}{\pi}\int_{0}^{\infty}\frac{\sqrt{\beta}dx}{\sqrt{x}(1+\beta
x)}\ln\left(\frac{N}{\alpha^{2}}.\frac{(1+x)^{N-1}\prod_{k=1}^{[\frac{N-1}{2}]}\mid
x-x_{k}^{B}\mid}{\prod_{k=1}^{[\frac{N}{2}]}\mid
x-x_{k}^{A}\mid^{3}}\right).$$
 Now making the following change of
variable $x=\frac{1}{\beta}\tan^{2}\frac{\Theta}{2}$, and taking
into account that degree of numerators and denominator are equal
for both even and odd values of N, we get
\newpage
$$h(\mu,\Phi_{N}(x,\alpha))=\frac{1}{\pi}\int_{0}^{\infty}d\theta\{\ln(\frac{N}{\alpha^{2}})+(N-1)\ln\mid\beta+1+(\beta-1)\cos\theta\mid
$$ $$ +\sum_{k=1}^{[\frac{N-1}{2}]}\ln
\mid1-x_{k}^{B}\beta+(1+x_{k}^{B}\beta)\cos\theta\mid
-3\sum_{k=1}^{[\frac{N}{2}]}\ln\mid
1-x_{k}^{A}\beta+(1+x_{k}^{A}\beta)\cos\theta\mid\}.$$ \\
 Using the following integrals:
$$ \frac{1}{\pi}\int_{0}^{\pi}\ln\mid a+b\cos\theta\mid= \left\{
\begin{array}{l}
\ln\mid\frac{a+\sqrt{a^{2}-b^{2}}}{2}\mid\quad\quad\mid a\mid
>\mid b\mid
\\ \ln\mid\frac{b}{2}\mid\quad\quad\quad\quad\quad\mid
a\mid \leq\mid b\mid,
\end{array}\right.$$
we get
\begin{equation}
 h(\mu,\Phi_{N}(x,\alpha))=\ln\left(\frac{N(1+\beta+2\sqrt{\beta})^{N-1}}{(\sum_{k=0}^{[ \frac{N}{2}]}C_{2k}^{N}\beta^{k})(\sum_{k=0}^{[
 \frac{N-1}{2}]}C_{2k+1}^{N}\beta^{k})}\right).
 \end{equation}
 \newpage
 \renewcommand{\thesection}{D}
\renewcommand{\theequation}{\thesection-\arabic{equation}}
\setcounter{equation}{0}
 {\large \appendix{\bf Appendix D}}: Derivation of entropy of coupled chaotic
 maps.
\\ The KS-entropy of $\Phi_{2,N_{2}}$ given in $(V-3)$ can be
written as
  \begin{equation}
h(\tilde{\mu},\tilde{\Phi}_{2,N_{2}})=
\int_{0}^{\infty}{\tilde{\mu} (x_{1})dx_{1}}
   \int_{0}^{\infty}{\tilde{\mu} (x_{2})dx_{2}}
   {(\ln{\mid \frac{\partial \tilde{X}_{2}}{\partial x_{1}}\mid}
   +\ln{\mid \frac{\partial \tilde{X}_{N_{2}}}{\partial
   x_{2}}\mid})}.
   \end{equation}
   Now, using the formulas $(III-5)$ the second integral reads
$$=\int_{0}^{\infty}{\tilde{\mu} (x_{1})dx_{1}}
   \int_{0}^{\infty}{\tilde{\mu} (x_{2})dx_{2}}
   \ln{\mid \frac{\partial}{\partial x_{1}}
   (\frac{1}{\alpha_{2}^{2}}.\tan^{2}(N_{2}\arctan\sqrt{x_{2}}))\mid}$$
or
    $$=\int_{0}^{\infty}{\tilde{\mu}(x_{1})dx_{1}}\int_{0}^{\infty}
                       {\tilde{\mu}(x_{2})dx_{2}}
                       \ln{\left( \mid \frac{N_{2}}{\alpha_{2}^{2}}.\frac{1}
                       {\sqrt{x_{2}}(1+x_{2})}
                      .\frac{\sin N_{2}(\arctan\sqrt{x_{2}})}
                           {\cos^{3}N_{2}(\arctan\sqrt{x_{2}})}\mid\right)},$$
where using the relation $(A-8)$, we get
\begin{equation}
  =\frac{1}{\pi}\int_{0}^{\infty}\frac{\sqrt{\beta_{2}}dx_{2}}{\sqrt{x_{2}}(1+\beta_{2}
    x_{2})}\ln\left(\frac{N_{2}}{\alpha_{2}^{2}}\mid\frac{(1+x_{2})^{N_{2}-1}B(-x_{2})}{x_{2}^{2}(A(-x_{2}))^{3}}\mid\right)
  \end{equation}
 Finally, calculating the above integral with the prescription of
Appendix C, we obtain:
 \begin{equation}
     =\ln\left(\frac{N_{2}(1+\beta_{2}+2\sqrt{\beta_{2}})^{N_{2}-1}}
     {(\sum_{k=0}^{[
     \frac{N_{2}}{2}]}C_{2k}^{N_{2}}\beta_{2}^{k})(\sum_{k=0}^{[
     \frac{N_{2}-1}{2}]}C_{2k+1}^{N_{2}}\beta_{2}^{k})}\right),
 \end{equation}
 Now, using the relation $(II-3)$ together with the invariance of
 measure, the first integral in $(D-1)$ can be
 $$=\frac{1}{\pi^{2}}\int_{0}^{\infty}\tilde{\mu}(x_{1})dx_{1}
                    \int_{0}^{\infty}\tilde{\mu}(x_{2})dx_{2}
                     \ln{\mid \frac{\partial }{\partial x_{1}}
                         \frac{\beta(x_{2})}{\beta(X_{2})}
                         .\frac{A_{2}^{2}(\frac{1}{\beta(x_{2})})}
                              {B_{2}^{2}(\frac{1}{\beta(x_{2})})}}$$
  $$\times \frac{\sin^{2}(2\arctan{\sqrt{x_{1}})}}
                {\cos^{2}(2\arctan{\sqrt{x_{1}})}}\mid $$
 \begin{equation}
 =\frac{1}{\pi^{2}}\int_{0}^{\infty}\tilde{\mu}(x_{1})dx_{1}
                    \int_{0}^{\infty}\tilde{\mu}(x_{2})dx_{2}
                    \ln{\mid \frac{\beta(x_{2})}{\beta(X_{2})}
                        .\frac{(1+\sqrt{\beta(x_{2})})^{2}}
                 {A_{2}(\beta(x_{2}))B_{2}(\beta(x_{2}))}\mid},
 \end{equation}
Due to the invivariance of measure the integral of the expression
$\frac{\beta(x_{2})}{\beta(X_{2})}$ vanishes, since according to
Appendix E, we have
\begin{equation}
\int_{0}^{\infty}dx\mu{(x)}\ln{(\beta(X_{2}))}
=\int_{0}^{\infty}dx\mu{(x)}\ln{(\beta(\Phi(x)))}
=\int_{0}^{\infty}dx\mu{(x)}\ln{(\beta(x_{2}))}\end{equation}
Making a change of variable
 $x=\frac{1}{\beta}\tan^{2}{\frac{\theta}{2}}$ in the first integral and
 omitting the expression $\frac{\beta(x_{2})}{\beta(X_{2})}$, it
 reduces to
    $$=\frac{2}{\pi}\int_{0}^{\frac{\pi}{2}}d\theta_{2}  {\left(
    \ln{((\beta_{2}(1+\beta_{1_{0}})+g)+(\beta_{2}(1+\beta_{1_{0}})+g)\cos\theta)})\right.}
    $$
    $$
    \left.\ln{(\beta_{2}+\beta_{2}\cos\theta)}+\ln{(\frac{3}{4}+\cos\theta+\frac{1}{4}\cos2\theta)}
    +\ln{(A+B\cos\theta+C\cos2\theta)}\right)$$
Which
$$ \left\{
\begin{array}{l}
A=\frac{1}{8}(3(1+\beta_{1_{0}})+\frac{2g\sqrt{\beta_{1_{0}}}}{\beta_{2}}+3(\frac{g}{\beta_{2}})^{2})
\\ B=\frac{1}{2}(1+\beta_{1_{0}}-(\frac{g}{\beta_{2}})^{2})  \\
C=\frac{1}{8}((1+\beta_{1_{0}})+\frac{2g\sqrt{\beta_{1_{0}}}}{\beta_{2}}+(\frac{g}{\beta_{2}})^{2}),
\end{array}\right.$$
 The above expression can be calculated by using the following
integral:
 $$\frac{1}{\pi}\int_{0}^{2\pi}d\theta\ln{(A+B\cos\theta+C\cos 2\theta)}=2\ln{\triangle}$$
\begin{equation}
\triangle
=\frac{1}{\pi}\left(\frac{\sqrt{A-3C+\sqrt{(A+C)^{2}-B^{2}}}}{2}+
                                \frac{\sqrt{A+B+C}-\sqrt{A-B+C}}{2}\right).
\end{equation}
where the above integral has been evaluated by using the well
known mean values theorem of analytic function
$$\frac{1}{\pi}\int_{0}^{2\pi}d\theta \ln{\mid
f(z_{0}+R\exp{i\theta})\mid}=\mid f(z_{0})\mid $$ by choosing
$f(z)=\alpha+\beta\exp{i\theta}+\gamma \exp{2i\theta}$ .
Therefore, the first integral in (D-1) reads
\newpage
 $$=\ln{2}+4\ln{\left(\sqrt{1+\sqrt{\beta_{1_{{0}}}}}+\sqrt{\frac{g}{\beta_{2}}}\right)}$$
         $$-2\ln{\left(\sqrt{1+\beta_{1_{0}}}+\frac{g}{\beta_{2}}+
         \sqrt{\frac{2g\sqrt{\beta_{1_{0}}}}{\beta_{2}}
         +\frac{2g}{\beta_{2}}\sqrt{1+\beta_{1_{0}}}}\right)}
   -\ln{2}$$
 \begin{equation}
=\ln{\frac{\left(\sqrt{1+\sqrt{\beta_{1_{0}}}}+\sqrt{\frac{g}{\beta_{2}}}\right)^{4}}{\left(\sqrt{1+\beta_{1_{0}}}+\frac{g}{\beta_{2}}+
    \sqrt{\frac{2g\sqrt{\beta_{1_{0}}}}{\beta_{2}}
    +\frac{2g}{\beta_{2}}\sqrt{1+\beta_{1_{0}}}}\right)^{2}}},
   \end{equation}
  Now, combing it with $(D-3)$ we get the expression (V-4).
  \newpage
  \renewcommand{\thesection}{E}
\renewcommand{\theequation}{\thesection-\arabic{equation}}
\setcounter{equation}{0}
 {\large \appendix{\bf Appendix E}}: Proof the integral identity
 $(D-5)$
\\ Denoting $\Phi(x)$ by y and inverting it we get
$$x_{k}=\Phi^{-1}(y),\quad k=1,2,...,n$$
 with n as the number of branch, we can write the lefthand side
of relation $(D-5)$ as
 \begin{equation}
\int_{0}^{\infty}dx\mu{(x)}f(\phi(x))=\sum_{k=1}^{n}\int_{(x_{i})_{k}}^{(x_{f})_{k}}dx_{k}\mu{(x_{k})}f(\Phi(x_{k})),
\end{equation}
where $(x_{i})_{k}$($(x_{f})_{k}$ denote the end coordinate of the
k-th branch) denote it initial coordinate of the k-th branch now,
making the change of variable for $x_{k}$ to y using the
relations, $\infty =\Phi{((x_{f})_{k})}$, $0=\Phi((x_{i})_{k})$
and $f(y)=f(\Phi(x_{k}))$, together with the FP equation we obtain
$$\int_{0}^{\infty}f(\Phi(x))dx=\int_{0}^{\infty}\left(\sum_{k=1}^{n}\mu{(x_{k}(y))}\frac{dx_{k}(y)}{dy}\right)$$

\begin{equation}
=\int_{0}^{\infty}dy\Phi(y)\mu{(y)}.
\end{equation}

\newpage
 {\bf Figures Captions}
 \\ Fig.1. Plot of $\Phi_{2}(x,\alpha)$ map for $\alpha=0.19$ (dashed curve)
           and plot of $\Phi_{3}(x,\alpha)$ map for $\alpha=0.19$, (solid
           curve). The location of maxima and minima and their values(0 or 1)
           are independent of parameter $\alpha $, as it is shown in the
           figure.
\\ Fig.4. Plot of $\Phi_{4}(x,\alpha)$ map for $\alpha=0.19$ (solid
          curve) and plot of $\Phi_{5}(x,\alpha)$ map for $\alpha=0.19$,
          (dashed curve). The location of maxima and minima and their
values(0 or 1) are independent of parameter $\alpha $, as it is
shown in the figure.
  \\ Fig.3. Bifurcation diagram of $\Phi_{2}(x,\alpha)$, for
$0<\alpha <2$, it is ergodic and for $0<\alpha <\infty$, it has
stable fixed point at $x=1$..
\\ Fig .4. Bifurcation diagram of $\Phi_{3}(x,\alpha)$, where  for
$\frac{1}{3}<\alpha <3$, it is ergodic and for $0<\alpha
<\frac{1}{3}$, it has stable fixed point at $x=0 $, while for
$3<\alpha <\infty$, it has stable fixed point $x=1$.
\\ Fig.5.Bifurcation diagram of $\Phi_{2,3}$, where for $0<\alpha_{1}<2$
and $\frac{1}{3}<\alpha_{2}< 3$ it is ergodic,  while for
$2<\alpha_{1}<\infty$ and  $0<\alpha_{2}<\frac{1}{3}$
 ($3<\alpha_{2}<\infty$) it has fixed point at $[0,0]$ ($[0,1]$).
 \\ Fig.6. Bifurcation diagram of $\Phi_{3,2}$, where for
 $\frac{1}{3}<\alpha_{1}<3$ and $0<\alpha_{2}<2$ it is ergodic ,
while for $0<\alpha_{1}<\frac{1}{3}$ ($3<\alpha_{1}<\infty)$ and
$0<\alpha_{2}<2$ ($2<\alpha_{2}<\infty$) it has fixed point at
$[0,0]$ ($[1,0]$).
\\ Fig.7. Bifurcation diagram of $\Phi_{3,3}$, where for
$\frac{1}{3}<\alpha_{1}<3$ and $\frac{1}{3}<\alpha_{2}<3$ it is
ergodic, while for $0<\alpha_{1}<\frac{1}{3}$
($3<\alpha_{1}<\infty$ ) and $0<\alpha_{2}<\frac{1}{3}$
($3<\alpha_{2}<\infty$ )it has fixed point at $[0,0]$ ($[1,1]$ ).
 \\ Fig.8. Solid curve shows the variation of
KS-entropy of $\Phi_{2}(x,\alpha)$, in terms of the parameter
$\alpha$ dots shows the variation of Lyapunov characteristic
exponent of $\Phi_{2}(x,\alpha)$, in terms of the parameter
$\alpha$.
\\Fig.9. Solid curve shows the variation of KS-entropy of
$\Phi_{3}(x,\alpha)$ in terms of the parameter $\alpha$, dots
shows the variation of Lyapunov characteristic exponent of
$\Phi_{3}(x,\alpha)$, in terms of the parameter $\alpha$. \\
Fig.10. Solid surface shows the variation of KS-entropy of
$\Phi_{2,2}$ in terms of the parameters $\alpha_1$ and $\alpha_2$,
while dotted surface shows the variation of Lyapunov
characteristic exponent of $\Phi_{2,2}$ in terms of the parameters
$\alpha_1$ and $\alpha_2$.
\\ Fig.11. Solid surface shows the
variation of KS-entropy of $\Phi_{4,4}$ in terms of the parameters
$\alpha_1$ and $\alpha_2$, while dotted surface shows the
variation of Lyapunov characteristic exponent of $\Phi_{4,4}$ in
terms of the parameters $\alpha_1$ and $\alpha_2$.
\end{document}